\documentclass[journal=jpclcd,manuscript=letter]{achemso}

\usepackage{chemformula} 
\usepackage[T1]{fontenc} 

\author{Philipp C. Schmid}
\author{Oskar Asvany}
\author{Thomas Salomon}
\author{Sven Thorwirth}
\author{Stephan Schlemmer}
\email{schlemmer@ph1.uni-koeln.de}
\affiliation[PH1]
{I.~Physikalisches Institut der Universit\"at zu K\"oln, K\"oln,
Germany}

\title[Leak-out Spectroscopy]
  {\includegraphics[scale=0.5]{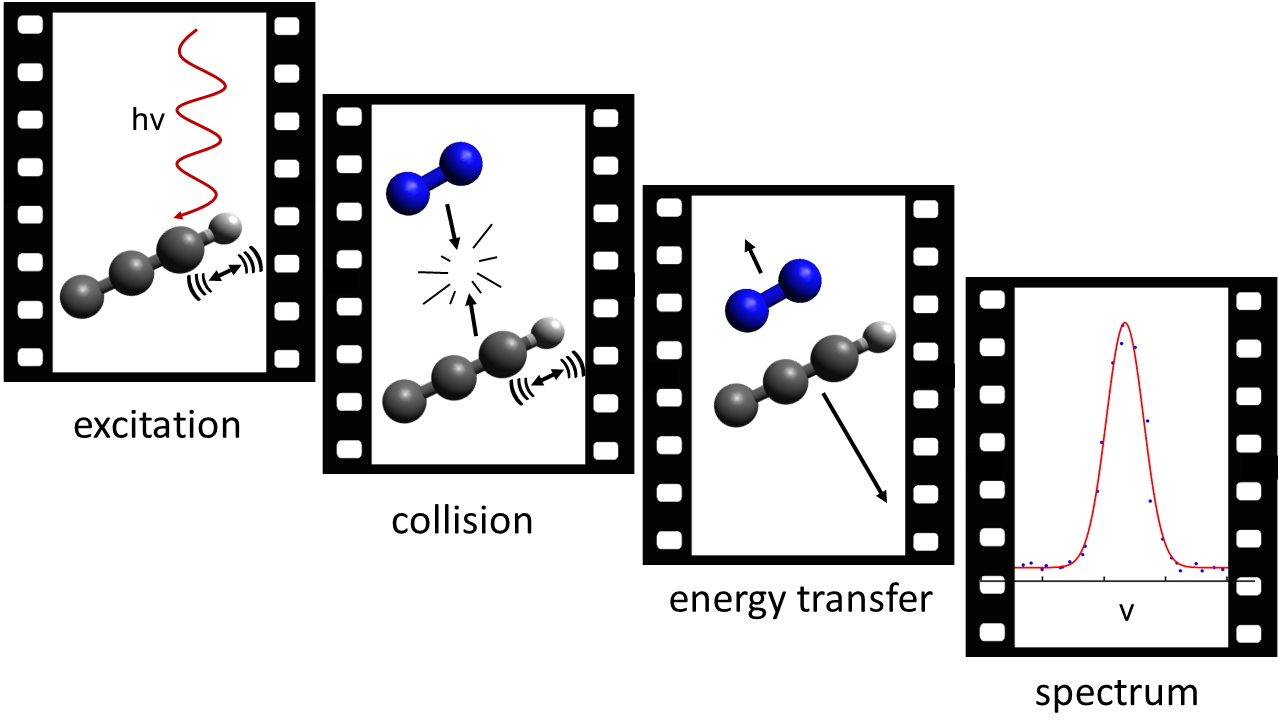}
  Leak-out Spectroscopy, a universal method of action spectroscopy in cold ion traps
  }

\abbreviations{IR}
\keywords{C$_3$H$^+$, ro-vibrational spectroscopy, ion traps, action spectroscopy, V-T-transfer, isomer selection}

\begin{document}

\begin{tocentry}
  \begin{center}
    \includegraphics[width=\linewidth]{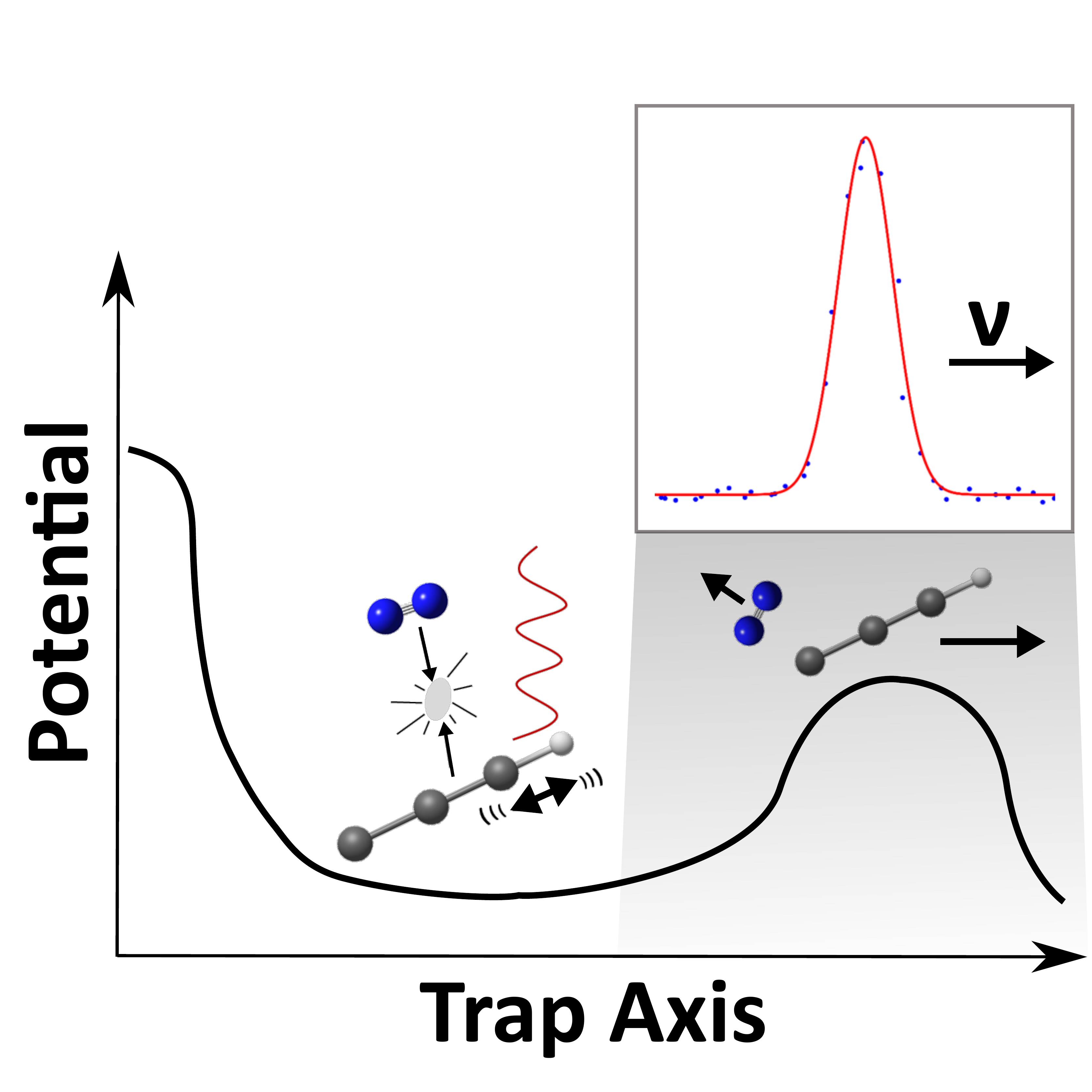}
  \end{center}
\end{tocentry}

\begin{abstract}
A novel method of spectroscopy in ion traps termed \textit{leak-out spectroscopy} (LOS) is presented.
Here, mass selected, cold ions are excited by an infrared laser. 
In a subsequent collision with a neutral buffer gas particle their internal energy is 
then transferred to kinetic energy. As a result, these ions {\it leak out} from the 
ion trap and are detected. 
The LOS scheme is generally applicable, very sensitive and close to background free when operated at low temperature. 
The potential of this method is demonstrated and 
characterized here for the first time by recording the rotationally 
resolved spectrum of the C-H stretching vibration $\nu_1$ of linear C$_3$H$^+$.
Besides performing high-resolution spectroscopy, 
this method opens up the way for analyzing the composition of trap content, e.g., determining isomer ratios,
by selectively expelling isomers or other isobaric ions from the trap.  
Likewise, LOS can be used to prepare clean samples of structural and nuclear spin isomers.   
\end{abstract}


\section{Introduction}
Over the last 35 years spectroscopy of molecular ions made tremendous progress on the basis of action spectroscopy, where the absorption of a photon is detected as a change of an ion signal. Following the pioneering work of Y.T.~Lee and coworkers \citep{oku85,oku88} especially infrared multiple photon dissociation (IRMPD) of mass selected ions and the photo-dissociation of loosely-bound ion-complexes, in particular in the form of messenger spectroscopy, 
became popular and very productive methods of ion spectroscopy. 
They have been applied to ions in molecular beams, ion guides and ion traps and 
even in commercial mass spectrometers \cite{Maitre:2007,GNS:2014,Oomens:2016,Riehn:2018}. 
For cases where the accuracy and resolution of IRMPD and messenger spectroscopy is not sufficient, 
alternative methods of action spectroscopy, e.g., various 
forms of Laser Induced Reactions (LIR) have been developed \citep{scl99,scl02}. 
Here, the products of a bi-molecular reaction are employed to detect the 
photo-absorption of the parent ion. This method requires appropriate 
reactions which are promoted by the internal excitation of the ion. 
A more general method, first developed by J. Maier and coworkers \citep{cha13}, 
uses the inhibition of complex formation (LIICG)\footnote{The method was termed 
Laser Induced Inhibition of Complex Growth, in short LIICG} of the parent ion with a rare gas atom, in particular with He. 
For this purpose cryogenic temperatures are required  such that ion-rare-gas clusters are formed and this signal is then altered by the internal excitation of the parent ion of interest. 

All these methods are predominantly used in tandem mass spectrometers 
where the parent ions are  mass selected by the first mass filter, 
then subjected to the exciting radiation in a cryogenic ion trap 
and the product ions are analyzed in the second mass filter 
before they are detected with near unit efficiency. 
With this repertoire of methods spectra of a large variety of 
molecular ions have been recorded. Even pure rotational spectra 
could be obtained by the use of double resonance schemes 
combining  millimeter wave excitation of rotational 
states with infrared lasers to create an action spectroscopic 
signal which is altered by the rotational excitation~\cite{asv21d}.
Despite the success story of action spectroscopy, 
many important ions could not be studied in high resolution so far,
the reason being, e.g., a missing adequate reaction for LIR, or 
ineffective complex formation for LIICG.

%
\begin{figure}
	\includegraphics[width=0.65\linewidth]{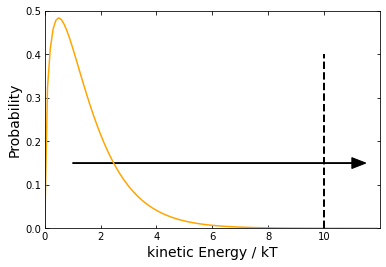}
    \caption{\label{V_T_transfer}
Thermal distribution of trapped ions. Ions which are accelerated by a vibration-to-translation 
energy transfer collision such that E$^{\prime}_{kin} >$~10~kT will leave the 
trap when the trap depth E$_A$ is of this order too. 
    }
\end{figure}
%

Here we report on a new, very universal approach of action spectroscopy, 
where the action relies on the inelastic energy transfer of an 
excited ion with a neutral collision partner. 
As a consequence, the ion gains kinetic energy and 
leaves the cold ion trap. The signal can then be detected as 
a loss of stored parent ions or as an arrival of escaping parent ions, 
even when the trap is nominally closed for cold ions. 
Based on the action at work we term this method \textit{leak-out spectroscopy} (LOS).
The idea of this method is depicted in Fig.~\ref{V_T_transfer} where a thermal Maxwell-Boltzmann distribution of the kinetic energy of trapped ions is displayed. The most probable kinetic energy is on the order of E$_{kin}$~=~kT and only very few ions are found with E$_{kin}$~$>$ 10~kT (see vertical line in Fig.\ref{V_T_transfer}). 
Based on this fact, the trap can be closed by applying an appropriate voltage to the exit electrode of the trap, resulting in a barrier (E$_{A}$) along the trap axis which exceeds this value, i.e., E$_{A}$~$>$ 10~kT. Thus the ions can be stored for very long times. 
If, however, an initially thermal ion is excited internally, 
e.g., to a vibrationally excited state,  a subsequent 
collision with a neutral collision partner
may transfer vibrational to translational energy (V-T-transfer).
This results in a change of kinetic energy $\Delta$E$_{kin}$, 
such that the kinetic energy after this collision can easily exceed the ion trap barrier height, E$_{A}$, 
as indicated by the horizontal arrow in Fig. \ref{V_T_transfer}. 
The accelerated ion can now escape the trap resulting in the 
desired action spectroscopy signal which only appears upon 
tuning the radiation source to resonant absorption. 

\begin{figure}[t]
	\includegraphics[width=0.65\linewidth]{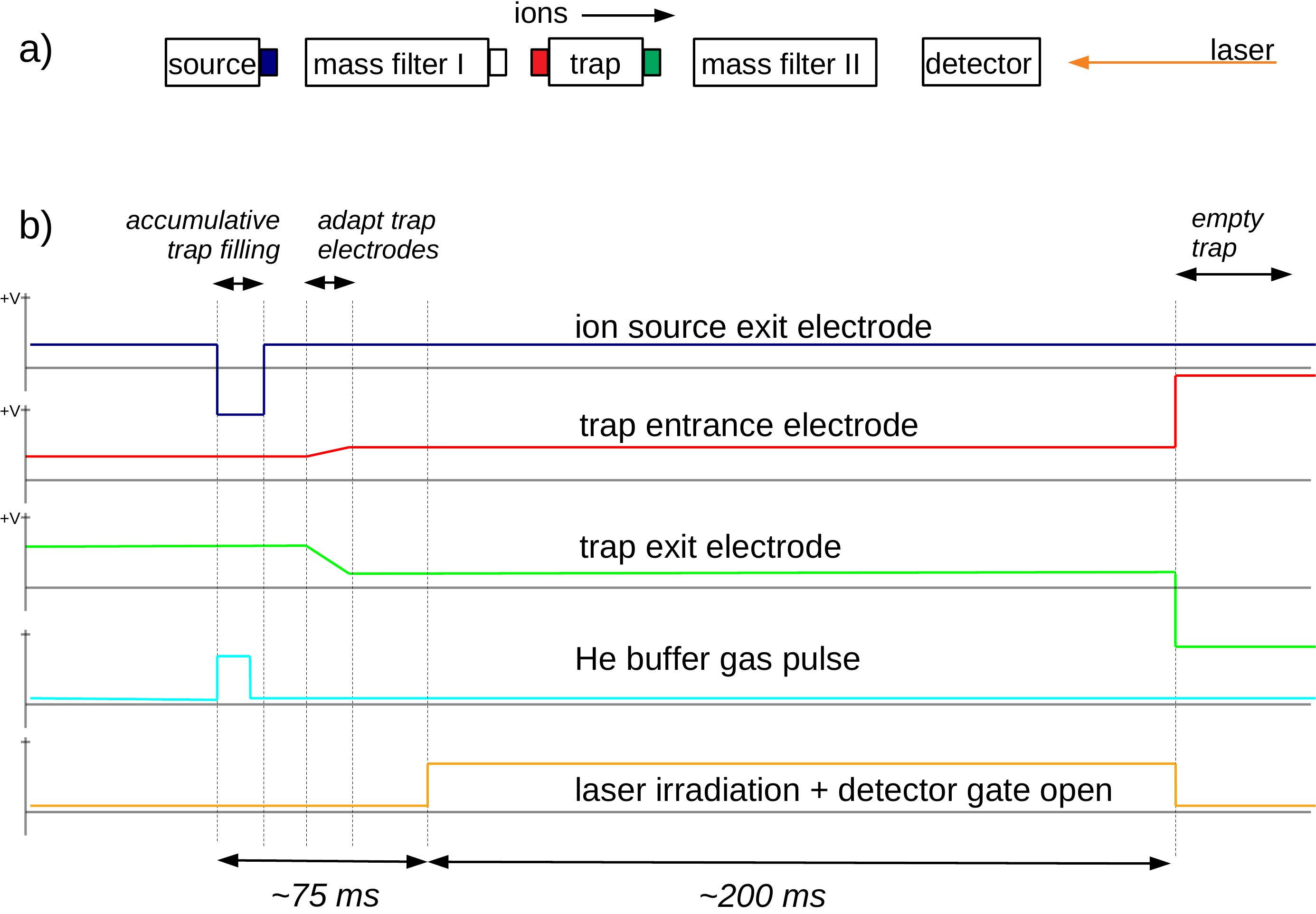}
    \caption{\label{sequence}
Schematic depiction (a) of the trap machine (LIRtrap) and 
(b) of the pulse sequence used for LOS (using the same color-coding as for the electrodes). Details are explained in the text.
    }
\end{figure}


\section{Experimental}
In order to prove this idea the cryogenic 22-pole ion trap LIRtrap was employed. 
It is the same instrument as already   used for the first LIR 
experiments \cite{scl99,scl02}.
Fig.~\ref{sequence}~a) shows schematically the tandem mass spectrometer with the central cryogenic 22-pole ion trap.

Ions are created in the storage ion \emph{source} (SIS) by electron bombardment. In the present case,  C$_3$H$^+$ and other small hydrocarbon ions are created from  allene precursor molecules (C$_3$H$_4$) which is admitted to the source at a pressure of about  10$^{-6}$~mbar. At the beginning of a trapping cycle
a train of ions is extracted from the source by pulsing the ion source exit electrodes as shown in 
the top blue trace of Fig.~\ref{sequence} b). 
These ions are mass selected in \emph{mass filter I} for mass 37 u (C$_3$H$^+$) 
which are then admitted to the \emph{trap} for several ten milliseconds termed \emph{accumulative trap filling}. 
During this time a pulse of He buffer gas (turqois trace in Fig.~\ref{sequence} b)) 
stops the ions in the potential of the 22-pole ion trap and
ensures thermalisation of ions to the cryogenic temperatures. 
At the end of this initial trapping phase the voltages of the 
entrance (red trace) and exit electrodes (green trace) of the \emph{trap} 
are adjusted to keep the stored ions but to allow fast ions to escape via the exit electrode as described in the introduction. 
During the following 200~ms the laser is admitted to the trap on the axis of the apparatus and the detector gate is opened to detect the fast ions leaving the trap (orange trace). 
For our first LOS experiments, the trap is constantly filled with N$_2$ gas at a
 density of about 10${^{12}}$/cm$^3$ and thus a collision rate of approximately 10${^3}$/s.

In order to avoid freezing of this gas,  the ion trap is held at about 40~K. Most collisions are elastic or change the rotational state of the molecules. But some collisions lead to the desired V-T-transfer described above and 
the ions leaving the trap upon an inelastic collision with the N$_2$ buffer gas are again mass analysed by \emph{mass filter II} as shown in Fig.~\ref{sequence}~a). 
In contrast to other forms of action spectroscopy where the mass of the stored ion is changed, here the second mass filter is also transmitting mass 37 u detecting the C$_3$H$^+$ ions leaving the trap. 
At the end of each trapping cycle any remaining ions are expelled from the trap to ensure that the next cycle starts with an empty trap. 
Before the next trapping cycle the laser frequency is stepped by about a tenth of the Doppler-width  (0.0005~cm$^{-1}$) of an absorption line and the experiment is repeated again. 
Instead of detecting the ions expelled from the trap during the trapping 
period also the ions remaining in the trap can be detected in another 
series of experiments such that the kinetics of the ion loss can be 
followed as a function of time as will be described below.

\section{Results}
As a first example for LOS, 
the C-H stretching vibration of the linear C$_3$H$^+$ (l-C$_3$H$^+$) 
molecular ion is excited by the intense light of a high-resolution infrared OPO system (Toptica Topo). 
As described above, the exit electrode of the trap is tuned in a way that the cold ions are barely held in the trap, 
i.e.\ some are constantly leaving the trap but with a very small rate only. When the exciting radiation is, 
however, in resonance with a ro-vibrational transition of  l-C$_3$H$^+$, a vibrationally 
inelastic collision leads to a considerable redistribution of the internal 
excitation (E$_{vib}~\sim~3170~$cm$^{-1}~\sim$~0.393~eV)  to the kinetic 
energy of both collision partners. In case of  l-C$_3$H$^+$ ($m = 37$~u) and N$_2$ ($m = 28$~u) 
the kinetic energy release is well shared between the 
two partners in the laboratory frame, 
with C$_3$H$^+$ taking a maximum of $\Delta$E$_{kin}$ = $28/(28+37) \cdot 393$~meV = 169~meV.
Now the fast C$_3$H$^+$ ions resulting from this  V-T-transfer
leave the trap across the E$_A$ barrier with a significantly enhanced rate.

\begin{figure}[t]
	\includegraphics[width=0.65\linewidth]{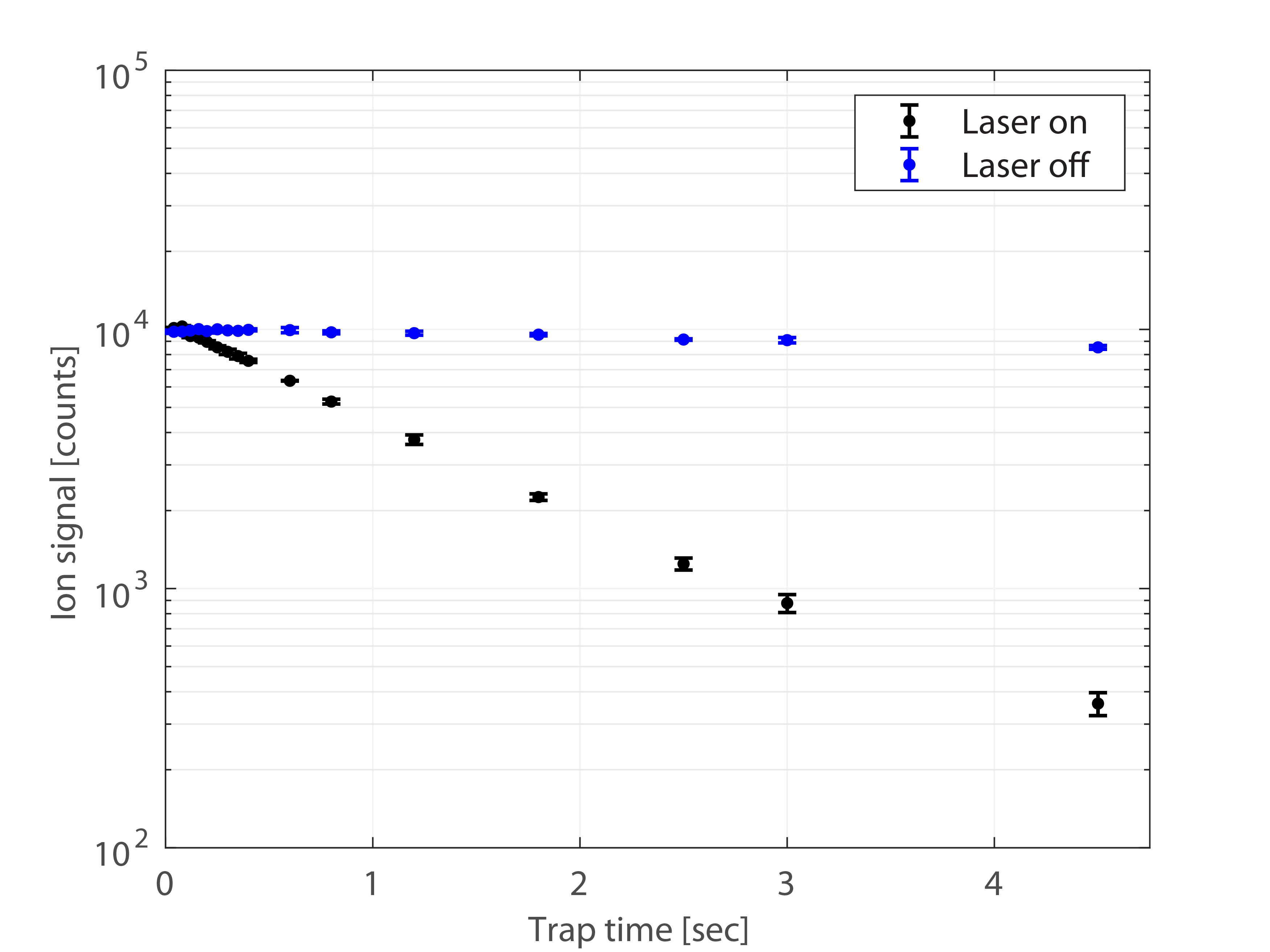}
    \caption{\label{ion-loss}
 Number of trapped l-C$_3$H$^+$ as a function of the trapping time. Less than 10~\% of the ions leave the trap without laser excitation. With the laser resonant to the R(5) transition the parent ions are lost with a 1/e time constant of approximately 1.5~s.   
    }
\end{figure}

\begin{figure}
	\includegraphics[width=\linewidth]{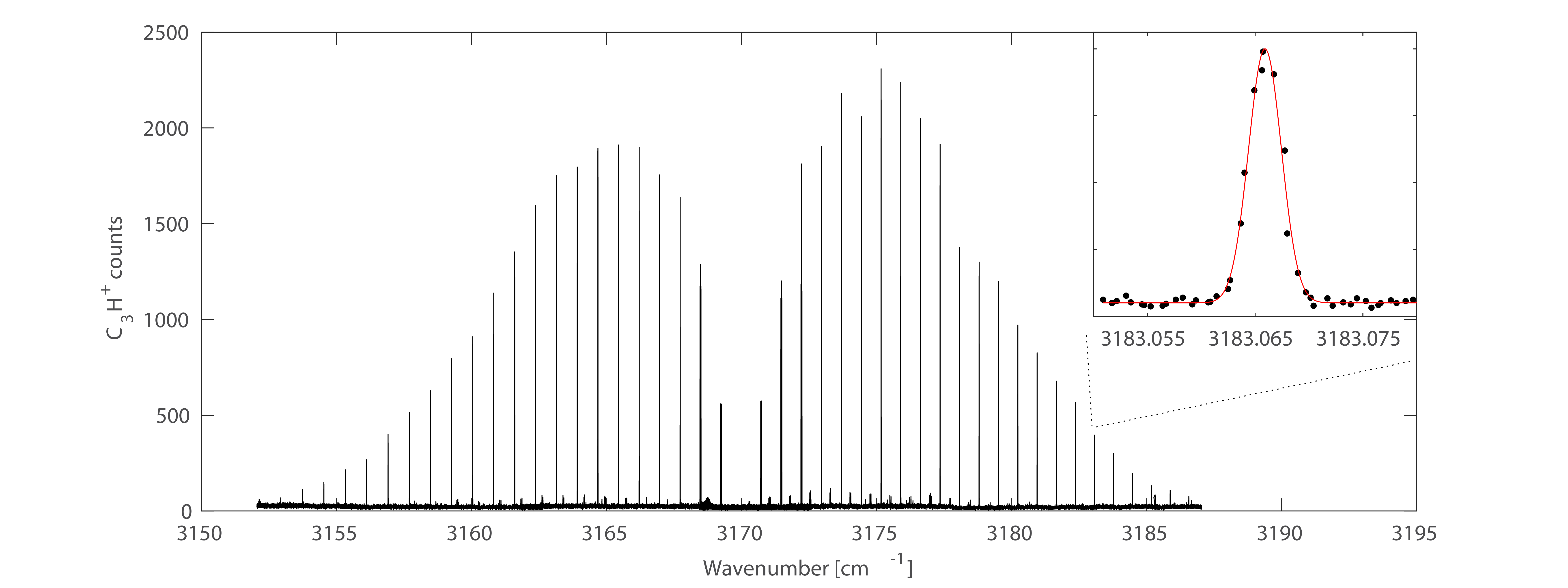}
    \caption{\label{infrared_spectrum}
Leak-out spectrum of the C-H stretching mode $\nu_1$ of l-C$_3$H$^+$, recorded by  
counting the number of escaping ions during the 200~ms lasting irradiation time. 
The $\nu_5$+$\nu_1$ $\leftarrow$ $\nu_5$ 
hot band is observed with an intensity of a few percent of the fundamental mode. The inset shows a zoom on the R(17) line fitted by a Gaussian profile corresponding to a kinetic temperature of T$_{kin}$ = 62(8)~K.
    }
\end{figure}

Fig.~\ref{ion-loss} shows the result of such an experiment as a function of the trapping time. 
Without laser excitation 90~\% of the about 10$^4$ parent ions 
are kept in the trap for more than 4 seconds. When the laser 
excites the R(5) transition of the C-H stretch of l-C$_3$H$^+$, 
the ion number drops exponentially to only a few percent after 4~s. 
This large enhancement of the ion loss rate can now easily be used 
to record the ro-vibrational spectrum of l-C$_3$H$^+$.
A spectrum recorded this way is given  in  Fig.~\ref{infrared_spectrum},
plotting the number of ions leaving the trap during the irradiation period as a function of the 
excitation wavenumber, measured by a wavemeter 
(Bristol Instruments, model 621 A IR, accuracy $\sim$0.001~cm$^{-1}$).
Only several ten ions leave the trap with the OPO 
off-resonant and the signal rises to more than 2000 ions on some 
lines of the R-branch.  Thus the recorded spectrum is almost background-free. 
The whole spectrum was measured in one scan which lasted only about 10 hours. 
The signal-to-noise ratio of this spectrum exceeds that of any action 
spectrum recorded in our laboratory by the aforementioned methods. 
It is in fact so high, that even a hot band of l-C$_3$H$^+$ is 
visible  with an intensity of a few percent 
of the main band as will be discussed below. 

%
%

The main spectrum shows a harmonic structure with a spacing of $2B\approx$0.75~cm$^{-1}$=22.5~GHz. 
It belongs to the linear C$_3$H$^+$ molecule for which the rotational constant of the 
ground vibrational state was determined with very high precision ($B=11244.950(2)$\,MHz) in an earlier experiment using the rotational-state-dependent 
attachment of He \cite{bru14}.
Fig.~\ref{infrared_spectrum} represents the first rotationally resolved vibrational 
spectrum of l-C$_3$H$^+$,   the $\nu_1$ C-H stretching band obtained in the present study. It exhibits a clear $P$- and $R$-branch 
with an intensity distribution which qualitatively follows 
that of an absorption spectrum of a linear species. 
For spectroscopic analysis the spectrum has been fitted to a 
standard linear molecule Hamiltonian.  
The molecular parameters derived from this analysis are given in Table~\ref{table:molecular:parameters}. 
Also the spectroscopic parameters of a weak hot band, identified as $\nu_1$+$\nu_5$ $\leftarrow$ $\nu_5$, 
and originating from the lowest energy C-C-C bending mode $\nu_5$, are given in this table. 
Detection of a $Q$-branch and $\ell$-type doubling in these weaker lines is a clear indication 
of the involvement of a degenerate bending vibration of the linear molecule.
For comparison, Table~\ref{table:molecular:parameters} also reports results
from previous experiment and high-level quantum-chemical calculations  
at the fc-CCSD(T)/ANO2 level
of theory. The latter were performed using the
CFOUR program \cite{cfour_JCP_2020} and complement other high level calculations reported in the literature
earlier (e.g., Refs \cite{botschwina_ApJ_787_72_2014,brunken_JPCA_123_8053_2019,schroeder_JMS_386_111628_2022}, and references therein).  

Overall good agreement between the experimental and calculated values is found. The new data may be reproduced to within experimental
accuracy with just a few molecular parameters, i.e., the lower and upper state rotational constants $B_v$ (alternatively, one rotational and 
a rotation-vibration interaction constant $\alpha_i$), centrifugal distortion terms,
an $\ell$-type doubling constant $q_i$ for the states involving the $\nu_5$ bending mode
as well as the vibrational band centers $\Tilde{\nu}_i$.
The $\nu_1$ vibrational wavenumber of untagged
C$_3$H$^+$ derived here is found blueshifted by some eight cm$^{-1}$ relative to C$_3$H$^+$--Ne studied earlier with infrared photodissociation \cite{brunken_JPCA_123_8053_2019}.
This finding is in qualitative accord with that in other
linear proton-bound weakly bound complexes (see, e.g., Ne--HCO$^+$, Ref. \cite{nizkorodov_JCP_105_1770_1996}) and in good agreement with a harmonic estimate of 13\,cm$^{-1}$
calculated at the CCSD(T)/aug-cc-pCVQZ level \cite{brunken_JPCA_123_8053_2019}.
Further information and details
on the spectroscopic analysis and calculations is given in the electronic supplementary material.

\begin{table}[ht]
    \centering
      \caption{Molecular parameters of l-C$_3$H$^+$ in selected vibrational states. 
      (in MHz unless noted otherwise). }
    \begin{tabular}{l r@{}l r@{}l r@{}l r@{}l }
              &   \multicolumn{2}{c}{$v$=0}  &   \multicolumn{2}{c}{$v_1=1$} &  \multicolumn{2}{c}{$v_5=1$} &  \multicolumn{2}{c}{$v_5=v_1=1$}\\  
\hline
    $\Tilde{\nu}_i$   / cm$^{-1}$  & & \multicolumn{2}{c}{3169.97769(16)$^a$}  &                   & &  \multicolumn{2}{l}{3168.76964(12)$^a$}& \\
    $\Tilde{\nu}_i$ / cm$^{-1}$  & &\multicolumn{2}{l}{3162.(2)$^b$}         &                   &  &  \multicolumn{2}{l}{3168.734$^c$} & \\
    $\chi_{ij}$/ cm$^{-1}$       & \multicolumn{4}{c}{$\cdots$}                                 &  \multicolumn{4}{c}{ $-$1.244$^d$/$-$1.20842(23)$^a$}  \\
    \\
    $B_v$               & 11244     &.9502(22)$^e$     & 11209  & .120(19)$^a$          & 11397&.90(33)$^a$            &   11362&.97(33)$^a$ \\
    $\alpha_i$          & \multicolumn{2}{c}{$\cdots$} & $-$35&.831(19)$^a$             & 152&.95(33)$^a$               & 118&.02(33)$^a$   \\  
    $\alpha_i$          & \multicolumn{2}{c}{$\cdots$} &    $-$32&.673$^d$              & 151&.892$^d$                  & 119&.219$^d$   \\  
    $D_v\times 10^3$    &     \multicolumn{4}{c}{7.710(75)$^f$ }                        &  \multicolumn{4}{c}{8.49(83)$^f$} \\
    $H_v\times 10^6$    & \multicolumn{4}{c}{0.76(28)$^f$}                              & \multicolumn{2}{c}{$\cdots$}    &   \multicolumn{2}{c}{$\cdots$} \\
    $q_i$               & \multicolumn{2}{c}{$\cdots$} & \multicolumn{2}{c}{$\cdots$}   & \multicolumn{4}{c}{68.348(96)$^f$}   \\
    $q_{i}$        & \multicolumn{2}{c}{$\cdots$} & \multicolumn{2}{c}{$\cdots$}   &\multicolumn{2}{c}{65.468$^d$}     &   \multicolumn{2}{c}{$\cdots$}                    \\

\hline
    \end{tabular}\\
    \raggedright
    $^a$ Experimental, this work.\\
    $^b$ From infrared photodissociation (IRPD) of the C$_3$H$^+$--Ne complex.\cite{brunken_JPCA_123_8053_2019} \\
    $^c$ From observed $\nu_1$ band center and calculated anharmonicity constant $\chi_{ij}$ (this work).\\
    $^d$ Calculated at the fc-CCSD(T)/ANO2 level, this work.\\    
    $^e$ Combined fit of $\nu_1$ band (this work) and pure rotational data (Ref. \cite{bru14}).\\
    $^f$ Common constant for both vibrational states, this work.
    \label{table:molecular:parameters}
\end{table}

Besides the spectral parameters additional information has been derived from the recorded spectrum.
The kinetic temperature of the trapped ions is determined from the Doppler profile of 
the spectral signal as shown in the inset in Fig.~\ref{infrared_spectrum} for the R(17) line to be $T_{kin} = 62(8)$~K which is somewhat elevated w.r.t.\ the trap wall temperature. 
Still, we conclude that 10$kT_{kin}$ = 54~meV 
is much smaller than  $\Delta$E$_{kin}$ = 169~meV as considered for the LOS-principle.
The rotational temperature  has been determined from a 
Boltzmann-plot of the line intensities, I$_\mathrm{J}$, of the spectrum shown 
in Fig.~\ref{infrared_spectrum} using the equation
\begin{equation}
ln \left( \frac{\rm{I}_J}{\rm{S}^{\rm{P,R}}_J}  \right) 
\sim 
-\frac{\rm{E}_J}{\rm{k} \rm{T}_{rot}}
\end{equation}
where S$^{\mathrm{P,R}}_\mathrm{J}$ is the H\"onl-London factor for a linear molecule and E$_\mathrm{J}$ is the 
rotational energy of the molecule in its vibrational ground state  as determined 
from the spectroscopic parameters given in Table~\ref{table:molecular:parameters}. 
The rotational temperature derived from this procedure is T$_{rot}$ = 50~K. This temperature is determined by the collisions of the
neutral gas at  T$_{neutral}$ = 40~K with the ions at  T$_{ion}$ = 62~K. The effective collision temperature is thus calculated to be 
T$_{col}$~=~(m$_{ion}$T$_{neutral}$+m$_{neutral}$T$_{ion}$)/(m$_{ion}$+m$_{neutral}$) =  49~K in very good agreement with the measured rotational temperature.

\section{Discussion}
This work presents a first example of LOS for C$_3$H$^+$, a molecular of astrophysical importance. High-quality spectroscopic parameters have been derived from the high-resolution ro-vibrational spectrum of this linear molecule.
On the basis of the findings in this work it is possible to discuss the 
principle of the LOS method, more specifically how the spectral contrast of LOS arises.
In the LOS process the vibrational energy of about E$_{vib}/hc$~=3170~cm$^{-1}$ 
is eventually distributed between the various degrees of freedom of both collision partners.
The V-T energy transfer is very inefficient due to the large 
energy mismatch between vibration and rotation. As a consequence only a small amount 
of energy is redistributed among the rotational states of the collision partners. 
The efficiency of a V-V energy transfer is also rather small by the same argument, 
i.e., there is a large mismatch between the various energetically allowed vibrational 
states of the two collision partners. Therefore, vibrational quenching is known 
to be a rather inefficient process altogether. For specific systems we and others 
found that only about one in 100 collisions lead to a V-T-transfer \cite{mar20}. 
Thus, quite a high number density of buffer gas is needed to promote this 
process as part of the LOS method. 
Although being unlikely, upon vibrational quenching a considerable 
fraction of the vibrational energy is thus transferred to translation. 
For molecules with many vibrational degrees of freedom quenching might leave the 
molecular ion in another vibrationally excited state which then yields 
less kinetic energy for the LOS process.

Preliminary studies on LOS of more complex ions such as protonated methanol or even protonated methane in our laboratory show that LOS also works perfectly for these molecules. For ions containing many more heavy elements it might be necessary to operate the trap at considerably lower temperatures to still observe a LOS signal. But at present we don't see a fundamental limit for the application of LOS for complex molecular ions.

Our considerations of the V-T-transfer process show that a considerable fraction 
of the vibrational energy $\Delta$E$_{vib}$ will be converted into translational energy, 
$\Delta$E$_{kin}$, with  $\Delta$E$_{vib} \approx$ $\Delta$E$_{kin}$ in the present case. 
For the success of LOS the molecular ion should take a considerable fraction of this energy. 
This can be achieved best for a heavy neutral collision partner, e.g., a heavy rare gas 
atom like Ar, Kr or Xe. In particular, for these partners energy  dissipation in internal 
degrees of freedom is excluded. 
But also N$_2$ will only transfer a small fraction of the energy 
into  molecular rotation as argued above. 
One more critical point for the method to work is that these neutral collision 
partners should not freeze to the walls of the trap. In that respect N$_2$ 
is a favorable choice for temperatures down to some 30-40~K. 
At lower temperatures condensable buffer gases will need to be supplied in a beam along the trap axis rather than as a background gas to avoid freezing. 
Another approach is to dilute these condensable gases in helium and admit them to the trap 
in a short pulse at the beginning of the trapping cycle. Using 3:1 mixtues of He:Ne or
He:Ar, we already succeeded to record high-resolution spectra of a handful of cations 
at a trap temperature of 4~K. The corresponding publications are in preparation.

As a result of these considerations we find that the collisions with the buffer gas are essential for LOS to work. The efficiency of the V-T-transfer process is only small compared to that of the thermalizing collisions, i.e. rotationally inelastic or elastic collisions, which prevent the ions from leaving the trap. Therefore, one finds a competing situation, where the V-T-transfer process is desired but subsequent thermalizing collisions are not. Thus the temporal behavior of the LOS process needs to be considered to find optimal conditions for LOS. 

The LOS sequence starts with the absorption of a photon, which we assume to be rather efficient, only limited by the spectral brightness of the  available radiation source.
Since the V-T-transfer should occur on time scales below a second, 
which is a typical trapping time, the number density $n$ of the buffer gas has to be on the order of 10$^{12}$~cm$^{-3}$.
With this density about 1000 ion-neutral collision occur per second,
and on average one to ten V-T-transfer collisions. 
Under these conditions, the fast ion produced after the V-T-transfer 
process collides about once 
per millisecond ($\tau~\sim$~1ms) and will eventually be thermalized, i.e., it can no longer leave the trap. 
Radiative decay also happens on  a similar time scale, i.e. $\tau_{rad}~\sim$~ms. Therefore, the fast ion has to leave the trap on this natural time scale, $\tau$, for LOS to work.

The chance to escape from the trap during this time increases with the velocity of 
the fast ion, $v$, and decreases with the linear dimension of the trap, $d$, which is of the order 0.1~$m$. 
Thus, the escape rate scales as $v/d$ and the escape probability as $v/d\cdot \tau$ because after the average collision 
time $\tau$ the ion is slowed down and trapped again for long. To increase the escape probability one would choose $\tau \sim$ $1/n$ long and thus the neutral gas density, $n$, low. However, a high gas density is needed to promote the V-T-transfer on the scale of typical trapping time. Therefore, in the experiments we find an optimum number density for the best LOS signal.

\begin{figure}
	\includegraphics[width=0.65\linewidth]{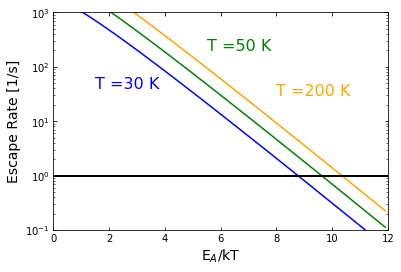}
    \caption{\label{Loss_rate}
Escape rate of ions from a thermal ensemble in a trap with a potential barrier of height 
E$_{A}$ in units of thermal energy $kT$ for three different trap temperatures. Escape rates below 1/s (horizontal line) ensure a safe long time trapping.
    }
\end{figure}

To evaluate the contrast of LOS the escape rate of the thermal ensemble of ions needs to be considered. Here $v$ is on the order of 100~$m/s$. Ions from the hot tail of 
the Maxwell-Boltzmann distribution may leave the trap when the escape barrier is chosen as 
E$_{A}$~$\approx$ 10~kT as described above. This fraction of fast ions leaving the trap, $P$, 
is given by integrating the Maxwell-Boltzmann distribution from E$_{kin}$=E$_{A}$ to $\infty$. 
In summary, we estimate the loss rate without laser excitation to be given by $R = v/d \cdot P$ which is plotted in 
Fig.~\ref{Loss_rate} as a function of the barrier height E$_{A}$ in units of the thermal energy $kT$.

%
%

It can be seen from this graph that the escape rate drops to less than 1/s (horizontal line in Fig.~\ref{Loss_rate}) for E$_{A}$ larger than 10~kT, which means the ions are kept a fraction of a second under these conditions. 
In fact, the off-resonance loss rate determined in Fig.~\ref{ion-loss} is considerably smaller, namely 2.3$\cdot$10$^{-2}$~s$^{-1}$. On-resonance with the  R(5) transition of l-C$_3$H$^+$, it increases by a factor of about 30 to 6.6$\cdot$10$^{-1}$~s$^{-1}$. Thus apparently we have chosen E$_{A}$ somewhat larger in the present experiments, but the process of LOS seems to be rather well described by the estimated escape rates on-resonance and off-resonance. 

From Fig.~\ref{Loss_rate} one reads that as long as E$_{A}$ is chosen in the right range, LOS should work also at room temperature (300~K). While this is true, the contrast between on- and off-resonance is given by the amount of energy transferred from vibrational to translational energy and how much this value lies above the thermal average $kT$. Therefore, LOS will work better under cryogenic conditions as demonstrated in this work.

In summary, we present a universal technique of action spectroscopy in cryogenic ion traps. 
The method relies on the inelastic collision with a buffer gas which should be chosen to promote 
the energy transfer. In the present case we used the vibrational excitation of the ion followed by the V-T-transfer. 
Our estimates show that the process of LOS is well described by deviations from a thermal 
ensemble in the trap. Other energetic processes like electronic excitations may 
be used as well to record a LOS signal. Also, the vibrational LOS introduced here can 
be combined with rotational spectroscopy. In this double resonance approach one 
would detect the change in the LOS signal when the rotational population of the 
ensemble of trapped ions is changed by a mm-wave radiation source as demonstrated 
earlier for other action spectroscopy methods~\cite{asv21d}. 
Moreover, LOS reveals the spectrum of the bare ion, 
C$_3$H$^+$ in the present case, and 
does not suffer from spectral shifts due to a tag as the comparison  with the spectra of the photodissociation (IRPD) of the C$_3$H$^+$--Ne complex shows. \cite{brunken_JPCA_123_8053_2019}
Therefore, LOS opens a whole suite 
of possibilities to record spectra of mass selected ions with an unprecedented 
sensitivity. 

Finally, similar to many methods of action spectroscopy 
in traps LOS also offers the chance to analyse 
the finite content of the trap. This is seen in Fig.~\ref{ion-loss} where 
on-resonance most of the ions leave the trap. 
This means that there is only one isomeric form of C$_3$H$^+$ in the trap, namely l-C$_3$H$^+$. 
In case a larger fraction of ions of a different species with the same nominal mass (m~=~37~u) would be co-trapped, 
these ions would not be addressed by the radiation and stay in the trap. This would be observed as a remaining number of ions in the trap in Fig.~\ref{ion-loss}. As such, LOS provides the possibility to determine the isomer composition of the initially trapped ensemble. Moreover, the escape of a specific isomer can be used to prepare a clean sample of selected isomers, conformers and nuclear spin isomers.
In a preliminary experiment on an ensemble of H$_3^+$ we were able to isolate the para or ortho fraction by leaking out the respective other fraction through resonant ro-vibrational excitation of the (J,K)~
=~(1,0) ortho- or (1,1) para-state, respectively.
In this sense, LOS is the beginning of new kind of experiments where reactant isomers can be selected and product isomer ratios can be analysed.
This we already showed for tagged ions in previous work \cite{jus19} but with LOS it works for bare ions.


\begin{acknowledgement}

This work has been supported via Collaborative Research Centre
956, sub-project B2, funded by the Deutsche
Forschungsgemeinschaft (DFG, project ID 184 018 867)
and through the ERC advanced grant (MissIons: 101020583). We also gratefully acknowledge support by DFG SCHL 341/15-1 (Ger\"atezentrum “Cologne Center for Terahertz Spectroscopy”).
\end{acknowledgement}

%
%
%

\bibliography{bib_arxiv}

\end{document}


\section{Quantum-chemical calculations}

In the present investigation, complementary quantum-chemical calculations were performed
at the coupled-cluster singles and doubles (CCSD) level 
augmented by a perturbative treatment of triple excitations, CCSD(T) \cite{raghavachari_chemphyslett_157_479_1989}, 
together with atomic natural orbital ANO2 basis set from
Alml\"of and Taylor \cite{almlof_JCP_86_4070_1987}. The frozen core approximation has been used throughout.
The equilibrium geometry was calculated using analytic gradient techniques
\cite{watts_chemphyslett_200_1-2_1_1992}, and is obtained as
$r_{\rm C1-C2}=1.344368$\,\AA , $r_{\rm C2-C3}=1.239122$\,\AA, and $r_{\rm C3-H}=1.080383$\,\AA .
Harmonic frequencies have been computed using analytic second-derivative techniques
\cite{gauss_chemphyslett_276_70_1997,stanton_IntRevPhysChem_19_61_2000}.
For anharmonic calculations, second-order vibrational perturbation theory (VPT2) \cite{mills_alphas}   
has been employed and additional numerical differentiation of analytic second derivatives has been applied 
to obtain the third and fourth derivatives required for the application of VPT2  \cite{stanton_IntRevPhysChem_19_61_2000,stanton_JCP_108_7190_1998}. The CCSD(T) method in combination
with ANO basis sets has been shown previously to provide vibrational wavenumbers of high quality \cite{mccaslin_MolPhys_111_1492_2013,thorwirth_JMS_251_220_2008}.
All calculations have been carried out using the CFOUR program package \cite{cfour_JCP_2020,harding_JChemTheoryComput_4_64_2008}. \\

The results of the force field calculations are summarized in Table \ref{qcc_results}
and they compare well with those presented in earlier high-level computations from
the literature. For the $\nu_1$ vibrational mode the present calculations yield a large
anharmonic correction resulting in a fundamental frequency lying some 60\,cm$^{-1}$
below the experimental value. Previous high-level calculations \cite{huang_ApJL_768_L25_2013,mladenovic_JCP_2014_224304_2014,brunken_JPCA_123_8053_2019}
have shown this mode to be affected by Fermi resonance and provide an anharmonic value
in much better agreement with the experimental value derived here. Using 
thresholds of $\Delta \omega = 50$\,cm$^{-1}$ and $\Phi_{ijk}=80$\,cm$^{-1}$
no resonance is observed at the present level of theory.

\begin{table}[h!]
\caption{fc-CCSD(T)/ANO2 vibrational wavenumbers, rotation-vibration interaction constants $\alpha_i$,
$\ell$-type doubling constants $q_i$ and band intensities of C$_3$H$^+$ (wavenumbers in cm$^{-1}$ ,
$\alpha_i$ and $q_i$ in MHz, band intensities in km/mol).\label{qcc_results}}
\begin{center}
\begin{tabular}{lrrrr|lr}
\hline \hline
Vib. & \multicolumn{2}{c}{Calc.$^a$}  & Exp.$^b$ & &  Para- & Calc.$^a$  \\ \cline{2-3}
 mode             & Harm. & Anharm.   &       &  Int.& meter        \\ \hline
$\nu_1(\sigma)$   & 3307  & 3108      & 3170  &  117 & $\alpha_1$             & 32.7  \\ 
$\nu_2(\sigma)$   & 2132  & 2087      & 2084  &  805 & $\alpha_2$             & 74.3  \\
$\nu_3(\sigma)$   & 1181  & 1185      & 1182  &   54 & $\alpha_3$             & 36.9  \\
$\nu_4(\pi)$      &  802  &  788      &  789  &   15 & $\alpha_4$/$q_{4}$     & $-$3.1/13.1 \\
$\nu_5(\pi)$      &  128  &  128      &  420  &    2 & $\alpha_5$/$q_{5}$     & $-$151.9/65.5 \\
 \hline
\end{tabular}
\end{center}
\end{table}

\section{Line list for C$_3$H$^+$}
\subsection{$\nu_1$ C-H stretching band of l-C$_3$H$^+$}
Truncated fit files from Pickett's SPFIT. Transitions are given in units of cm$^{-1}$ and molecular parameters in units of MHz.
Lines with residuals (DIFF.) larger than three times the experimental uncertainty have been omitted from the fits.
\begin{scriptsize}
\begin{verbatim}
                                              EXP.FREQ.  -  CALC.FREQ.  - DIFF.     - EXP.ERR.- EST.ERR.
    1:   2  0  1  0                           44979.53600   44979.55402   -0.01802    0.01400   0.00678
    2:   3  0  2  0                           67468.87440   67468.86932    0.00508    0.00850   0.00693
    3:   4  0  3  0                           89957.63700   89957.63210    0.00490    0.01300   0.00759
    4:   5  0  4  0                          112445.64900  112445.66071   -0.01171    0.02000   0.01735
    5:   1  1  0  0                          3170.7252000  3170.7254760 -0.0002760 -0.0007000 0.0001560
    6:   2  1  1  0                          3171.4706000  3171.4708704 -0.0002704 -0.0007000 0.0001541
    7:   0  1  1  0                          3169.2271000  3169.2275037 -0.0004037 -0.0007000 0.0001569
    8:   3  1  2  0                          3172.2136000  3172.2138621 -0.0002621 -0.0007000 0.0001513
    9:   1  1  2  0                          3168.4751000  3168.4749383  0.0001617 -0.0007000 0.0001560
   10:   4  1  3  0                          3172.9544000  3172.9544451 -0.0000451 -0.0007000 0.0001477
   11:   2  1  3  0                          3167.7204000  3167.7199947  0.0004053 -0.0007000 0.0001541
   12:   3  1  4  0                          3166.9629000  3166.9626793  0.0002207 -0.0007000 0.0001513
   13:   5  1  4  0                          3173.6928000  3173.6926132  0.0001868 -0.0007000 0.0001433
   14:   4  1  5  0                          3166.2031000  3166.2029979  0.0001021 -0.0007000 0.0001477
   15:   6  1  5  0                          3174.4283000  3174.4283604 -0.0000604 -0.0007000 0.0001381
   16:   7  1  6  0                          3175.1619000  3175.1616810  0.0002190 -0.0007000 0.0001325
   17:   5  1  6  0                          3165.4410000  3165.4409567  0.0000433 -0.0007000 0.0001432
   18:   8  1  7  0                          3175.8926000  3175.8925690  0.0000310 -0.0007000 0.0001265
   19:   6  1  7  0                          3164.6766000  3164.6765615  0.0000385 -0.0007000 0.0001381
   20:   9  1  8  0                          3176.6209000  3176.6210188 -0.0001188 -0.0007000 0.0001204
   21:   7  1  8  0                          3163.9099000  3163.9098181  0.0000819 -0.0007000 0.0001324
   22:  10  1  9  0                          3177.3470000  3177.3470250 -0.0000250 -0.0007000 0.0001146
   23:   8  1  9  0                          3163.1407000  3163.1407322 -0.0000322 -0.0007000 0.0001264
   24:  11  1 10  0                          3178.0708000  3178.0705821  0.0002179 -0.0007000 0.0001097
   25:   9  1 10  0                          3162.3694000  3162.3693092  0.0000908 -0.0007000 0.0001203
   26:  12  1 11  0                          3178.7920000  3178.7916850  0.0003150 -0.0007000 0.0001063
   27:  10  1 11  0                          3161.5954000  3161.5955546 -0.0001546 -0.0007000 0.0001145
   28:  13  1 12  0                          3179.5105000  3179.5103287  0.0001713 -0.0007000 0.0001052
   29:  11  1 12  0                          3160.8191000  3160.8194734 -0.0003734 -0.0007000 0.0001098
   30:  14  1 13  0                          3180.2269000  3180.2265085  0.0003915 -0.0007000 0.0001074
   31:  12  1 13  0                          3160.0407000  3160.0410707 -0.0003707 -0.0007000 0.0001068
   32:  13  1 14  0                          3159.2600000  3159.2603512 -0.0003512 -0.0007000 0.0001067
   33:  15  1 14  0                          3180.9406000  3180.9402199  0.0003801 -0.0007000 0.0001137
   34:  16  1 15  0                          3181.6517000  3181.6514586  0.0002414 -0.0007000 0.0001248
   35:  14  1 15  0                          3158.4770000  3158.4773194 -0.0003194 -0.0007000 0.0001107
   36:  17  1 16  0                          3182.3604000  3182.3602205  0.0001795 -0.0007000 0.0001412
   37:  15  1 16  0                          3157.6918000  3157.6919797 -0.0001797 -0.0007000 0.0001200
   38:  18  1 17  0                          3183.0667000  3183.0665021  0.0001979 -0.0007000 0.0001634
   39:  16  1 17  0                          3156.9044000  3156.9043358  0.0000642 -0.0007000 0.0001355
   40:  19  1 18  0                          3183.7705000  3183.7702999  0.0002001 -0.0007000 0.0001919
   41:  17  1 18  0                          3156.1143000  3156.1143917 -0.0000917 -0.0007000 0.0001578
   42:  20  1 19  0                          3184.4717000  3184.4716108  0.0000892 -0.0007000 0.0002274
   43:  18  1 19  0                          3155.3219000  3155.3221506 -0.0002506 -0.0007000 0.0001873
   44:  21  1 20  0                          3185.1705000  3185.1704322  0.0000678 -0.0007000 0.0002708
   45:  19  1 20  0                          3154.5273000  3154.5276157 -0.0003157 -0.0007000 0.0002247
   46:  22  1 21  0                          3185.8667000  3185.8667617 -0.0000617 -0.0007000 0.0003234
   47:  20  1 21  0                          3153.7307000  3153.7307896 -0.0000896 -0.0007000 0.0002708
   48:  23  1 22  0                          3186.5602000  3186.5605973 -0.0003973 -0.0007000 0.0003866
   49:  21  1 22  0                          3152.9319000  3152.9316747  0.0002253 -0.0007000 0.0003266
   50:  22  1 23  0                          3152.1304000  3152.1302729  0.0001271 -0.0007000 0.0003936
NORMALIZED DIAGONAL:
    1  1.00000E+000    2  1.00000E+000    3  6.57083E-001    4  3.82678E-001    5  9.80873E-001
MARQUARDT PARAMETER = 0, TRUST EXPANSION = 1.00
                                NEW PARAMETER (EST. ERROR) -- CHANGE THIS ITERATION
   1            11          E       95033540.2( 47)             -0.0    
   2           199         B0      11244.95015(221)         -0.00000    
   3           111      alpha         -35.8306(188)           0.0000    
   4           299        -D0           -7.710( 75)E-03        0.000E-03
   5           399          H            0.756(276)E-06       -0.000E-06
 MICROWAVE AVG =       -0.004938 MHz, IR AVG =        0.00000
 MICROWAVE RMS =        0.011310 MHz, IR RMS =        0.00023
 END OF ITERATION  2 OLD, NEW RMS ERROR=        0.38357         0.38357
\end{verbatim}
\end{scriptsize}

\subsection{$\nu_1+\nu_5-\nu_5$ hot-band of l-C$_3$H$^+$}
Truncated fit files from Pickett's SPFIT. Transitions are given in units of cm$^{-1}$ and molecular parameters in units of MHz.
Lines with residuals (DIFF.) larger than three times the experimental uncertainty have been omitted from the fits. The ground state rotational constant of the previous fit has been used to determine the rotation-vibration coupling constants of the involved vibrational states. 
\begin{scriptsize}

\begin{verbatim}
                                            EXP.FREQ.    - CALC.FREQ.  - DIFF.     - EXP.ERR.- EST.ERR.-
    1:   2 -1  2  1  1  1                    3170.2781000  3170.2788508 -0.0007508 -0.0007000 0.0001240
    2:   1 -1  2  1  1  1                    3168.7720000  3168.7718673  0.0001327 -0.0007000 0.0001199
    3:   1  1  2  1 -1  1                    3168.7628000  3168.7627479  0.0000521 -0.0007000 0.0001204
    4:   2  1  2  1 -1  1                    3170.2877000  3170.2879702 -0.0002702 -0.0007000 0.0001236
    5:   1 -1  2  2  1  1                    3167.2416000  3167.2419849 -0.0003849 -0.0007000 0.0001291
    6:   2 -1  2  2  1  1                    3168.7495000  3168.7489684  0.0005316 -0.0007000 0.0001195
    7:   3 -1  2  2  1  1                    3171.0423000  3171.0436243 -0.0013243 -0.0007000 0.0001280
    8:   2  1  2  2 -1  1                    3168.7768000  3168.7763265  0.0004735 -0.0007000 0.0001180
    9:   1  1  2  2 -1  1                    3167.2508000  3167.2511043 -0.0003043 -0.0007000 0.0001285
 ***** NEXT LINE NOT USED IN FIT
   10:   3  1  2  2 -1  1                    3171.0240000  3171.0299452 -0.0059452 -0.0007000 0.0001286
   11:   2 -1  2  3  1  1                    3166.4881000  3166.4883595 -0.0002595 -0.0007000 0.0001344
   12:   4 -1  2  3  1  1                    3171.7778000  3171.7786892 -0.0008892 -0.0007000 0.0001345
   13:   3 -1  2  3  1  1                    3168.7822000  3168.7830154 -0.0008154 -0.0007000 0.0001178
   14:   2  1  2  3 -1  1                    3166.4748000  3166.4746804  0.0001196 -0.0007000 0.0001353
   15:   4  1  2  3 -1  1                    3171.7964000  3171.7969280 -0.0005280 -0.0007000 0.0001337
 ***** NEXT LINE NOT USED IN FIT
   16:   3  1  2  3 -1  1                    3168.7233000  3168.7282991 -0.0049991 -0.0007000 0.0001211
   17:   5 -1  2  4  1  1                    3172.5477000  3172.5478744 -0.0001744 -0.0007000 0.0001398
   18:   4 -1  2  4  1  1                    3168.7005000  3168.7007400 -0.0002400 -0.0007000 0.0001285
   19:   3 -1  2  4  1  1                    3165.7041000  3165.7050662 -0.0009662 -0.0007000 0.0001420
 ***** NEXT LINE NOT USED IN FIT
   20:   3  1  2  4 -1  1                    3165.7177000  3165.7233050 -0.0056050 -0.0007000 0.0001407
   21:   5  1  2  4 -1  1                    3172.5246000  3172.5250759 -0.0004759 -0.0007000 0.0001408
   22:   4  1  2  4 -1  1                    3168.7918000  3168.7919339 -0.0001339 -0.0007000 0.0001225
   23:   6 -1  2  5  1  1                    3173.2692000  3173.2690986  0.0001014 -0.0007000 0.0001466
   24:   4 -1  2  5  1  1                    3164.9554000  3164.9559476 -0.0005476 -0.0007000 0.0001466
   25:   5 -1  2  5  1  1                    3168.8031000  3168.8030821  0.0000179 -0.0007000 0.0001354
   26:   4  1  2  5 -1  1                    3164.9325000  3164.9331491 -0.0006491 -0.0007000 0.0001484
   27:   5  1  2  5 -1  1                    3168.6665000  3168.6662911  0.0002089 -0.0007000 0.0001447
   28:   6  1  2  5 -1  1                    3173.2966000  3173.2964568  0.0001432 -0.0007000 0.0001453
   29:   6 -1  2  6  1  1                    3168.6253000  3168.6249525  0.0003475 -0.0007000 0.0001714
   30:   7 -1  2  6  1  1                    3174.0427000  3174.0426683  0.0000317 -0.0007000 0.0001495
   31:   5 -1  2  6  1  1                    3164.1589000  3164.1589360 -0.0000360 -0.0007000 0.0001537
   32:   5  1  2  6 -1  1                    3164.1856000  3164.1862942 -0.0006942 -0.0007000 0.0001513
   33:   6  1  2  6 -1  1                    3168.8168000  3168.8164599  0.0003401 -0.0007000 0.0001585
   34:   7  1  2  6 -1  1                    3174.0109000  3174.0107504  0.0001496 -0.0007000 0.0001513
   35:   6 -1  2  7  1  1                    3163.4141000  3163.4143515 -0.0002515 -0.0007000 0.0001544
   36:   8 -1  2  7  1  1                    3174.7503000  3174.7500245  0.0002755 -0.0007000 0.0001543
   37:   7 -1  2  7  1  1                    3168.8324000  3168.8320673  0.0003327 -0.0007000 0.0001921
   38:   8  1  2  7 -1  1                    3174.7874000  3174.7865021  0.0008979 -0.0007000 0.0001520
   39:   7  1  2  7 -1  1                    3168.5774000  3168.5767242  0.0006758 -0.0007000 0.0002086
   40:   6  1  2  7 -1  1                    3163.3824000  3163.3824336 -0.0000336 -0.0007000 0.0001576
   41:   8 -1  2  8  1  1                    3168.5231000  3168.5216060  0.0014940 -0.0007000 0.0002556
   42:   7 -1  2  8  1  1                    3162.6052000  3162.6036488  0.0015512 -0.0007000 0.0001598
   43:   9 -1  2  8  1  1                    3175.5292000  3175.5279514  0.0012486 -0.0007000 0.0001526
   44:   9  1  2  8 -1  1                    3175.4882000  3175.4869142  0.0012858 -0.0007000 0.0001557
   45:   8  1  2  8 -1  1                    3168.8502000  3168.8499043  0.0002957 -0.0007000 0.0002356
   46:   7  1  2  8 -1  1                    3162.6398000  3162.6401263 -0.0003263 -0.0007000 0.0001556
   47:  10 -1  2  9  1  1                    3176.2196000  3176.2214125 -0.0018125 -0.0007000 0.0001556
   48:   9 -1  2  9  1  1                    3168.8706000  3168.8699709  0.0006291 -0.0007000 0.0002880
   49:   8 -1  2  9  1  1                    3161.8635000  3161.8636255 -0.0001255 -0.0007000 0.0001554
   50:  10  1  2  9 -1  1                    3176.2688000  3176.2670095  0.0017905 -0.0007000 0.0001516
   51:   8  1  2  9 -1  1                    3161.8234000  3161.8225882  0.0008118 -0.0007000 0.0001608
 ***** NEXT LINE NOT USED IN FIT
   52:   9  1  2  9 -1  1                    3168.4636000  3168.4595981  0.0040019 -0.0007000 0.0003117
   53:   9 -1  2 10  1  1                    3161.0405000  3161.0392588  0.0012412 -0.0007000 0.0001615
 ***** NEXT LINE NOT USED IN FIT
   54:  11 -1  2 10  1  1                    3176.9904000  3177.0036696 -0.0132696 -0.0007000 0.0001501
   55:  10  1  2 10 -1  1                    3168.8939000  3168.8922672  0.0016328 -0.0007000 0.0003486
   56:  11  1  2 10 -1  1                    3176.9526000  3176.9535129 -0.0009129 -0.0007000 0.0001553
   57:   9  1  2 10 -1  1                    3161.0859000  3161.0848558  0.0010442 -0.0007000 0.0001547
 ***** NEXT LINE NOT USED IN FIT
   58:  10 -1  2 11  1  1                    3160.3014000  3160.3038240 -0.0024240 -0.0007000 0.0001555
   59:  12 -1  2 11  1  1                    3177.6829000  3177.6832083 -0.0003083 -0.0007000 0.0001567
   60:  12  1  2 11 -1  1                    3177.7366000  3177.7379247 -0.0013247 -0.0007000 0.0001501
 ***** NEXT LINE NOT USED IN FIT
   61:  10  1  2 11 -1  1                    3160.2561000  3160.2536673  0.0024327 -0.0007000 0.0001639
 ***** NEXT LINE NOT USED IN FIT
   62:  11 -1  2 12  1  1                    3159.4532000  3159.4658205 -0.0126205 -0.0007000 0.0001709
   63:  13 -1  2 12  1  1                    3178.4695000  3178.4697682 -0.0002682 -0.0007000 0.0001551
   64:  11  1  2 12 -1  1                    3159.5186000  3159.5205369 -0.0019369 -0.0007000 0.0001610
   65:  13  1  2 12 -1  1                    3178.4103000  3178.4104921 -0.0001921 -0.0007000 0.0001630
   66:  14 -1  2 13  1  1                    3179.1353000  3179.1353575 -0.0000575 -0.0007000 0.0001782
   67:  12 -1  2 13  1  1                    3158.7345000  3158.7350014 -0.0005014 -0.0007000 0.0001752
   68:  14  1  2 13 -1  1                    3179.1993000  3179.1991933  0.0001067 -0.0007000 0.0001692
   69:  12  1  2 13 -1  1                    3158.6754000  3158.6757253 -0.0003253 -0.0007000 0.0001861
   70:  15 -1  2 14  1  1                    3179.9259000  3179.9261930 -0.0002930 -0.0007000 0.0001959
   71:  13 -1  2 14  1  1                    3157.8845000  3157.8833883  0.0011117 -0.0007000 0.0002128
   72:  13  1  2 14 -1  1                    3157.9466000  3157.9472241 -0.0006241 -0.0007000 0.0002014
   73:  15  1  2 14 -1  1                    3179.8576000  3179.8577976 -0.0001976 -0.0007000 0.0002055
   74:  16 -1  2 15  1  1                    3180.5785000  3180.5778056  0.0006944 -0.0007000 0.0002467
   75:  14 -1  2 15  1  1                    3157.1560000  3157.1572120 -0.0012120 -0.0007000 0.0002416
   76:  16  1  2 15 -1  1                    3180.6514000  3180.6507608  0.0006392 -0.0007000 0.0002371
   77:  14  1  2 15 -1  1                    3157.0905000  3157.0888165  0.0016835 -0.0007000 0.0002529
   78:  17 -1  2 16  1  1                    3181.3729000  3181.3728896  0.0000104 -0.0007000 0.0002929
   79:  15  1  2 16 -1  1                    3156.3640000  3156.3649718 -0.0009718 -0.0007000 0.0002962
   80:  16 -1  2 17  1  1                    3155.5704000  3155.5705103 -0.0001103 -0.0007000 0.0003652
   81:  18  1  2 17 -1  1                    3182.0917000  3182.0925728 -0.0008728 -0.0007000 0.0003630
    8 Lines rejected from fit
NORMALIZED DIAGONAL:
    1  1.00000E+000    2  1.00000E+000    3  1.00000E+000    4  6.72297E-001    5  5.82020E-002    6  9.95271E-001
    7  9.88957E-001
MARQUARDT PARAMETER = 0, TRUST EXPANSION = 1.00
                                NEW PARAMETER (EST. ERROR) -- CHANGE THIS ITERATION
   1            11          E              0.0(100)E-36          0.0E-36
   2            22          E       94997323.9( 36)              0.0    
   3           199         B0  11244.950148714(  0)     -0.000000000    
   4           111      alpha           152.95( 33)            -0.00    
   5           122      alpha           118.02( 33)            -0.00    
   6           299        -D0            -8.49( 83)E-03         0.00E-03
   7         40099        q/2           34.174( 48)           -0.000    
 MICROWAVE AVG =        0.000000 MHz, IR AVG =        0.00000
 MICROWAVE RMS =        0.000000 MHz, IR RMS =        0.00079
 END OF ITERATION  7 OLD, NEW RMS ERROR=        1.13124         1.13124
\end{verbatim}
\end{scriptsize}

\bibliography{LIRTRAP,sthorwirth_bibdesk}